# Trusted Authentication using hybrid security algorithm in VANET

Prasanna Venkatesan E, Kristen Titus W

Department of Software Engineering,
SRM Institute of Science and Technology, Chennai

**ABSTRACT**:

Vehicular Ad Hoc Networks (VANETs) improves traffic management and reduce the amount of road accidents by providing safety applications. However, VANETs are vulnerable to variety of security attacks from malicious entities. An authentication is an integral a neighborhood of trust establishment and secure communications between vehicles. The Road-side Unit (RSU) evaluates trust-value and the Agent Trusted Authority (ATA) helps in computing the trust-value of auto supported its reward-points. The communication between nodes is enhanced, this can reduce 50% of road accidents. The security of the VANET is improved. We propose the utilization of Elliptic Curve Cryptography in the design of an efficient data encryption/decryption system for sensor nodes in a wireless network. Elliptic Curve Cryptography can provide impressive levels of security standards while keeping down the cost of certain issues, primarily storage space. Sensors will benefit from having to store relatively smaller keys coupled with increased computational capability and this will be a stronger design as the bit-level security is improved. Thus, reducing the time delay between the nodes and to provide better results between them we have made use of this method. The implementation of this work is done with NS2 software.

**INTRODUCTION**:

The vehicular ad hoc network (VANET) has become most popular state-of-the-art technology in the domain of intelligent transportation systems. A vehicle in VANET is embedded with an on-board unit (OBU) which enables vehicle to-vehicle (V2V) communications and vehicle-to-infrastructure (V2I) communications through RSUs. Elliptic Key Cryptography, a public key cryptography scheme for secure localization and authentication nodes is proposed. The key exchange between the nodes is done by using unique ECC algorithm. Mostly the main reason for attraction of ECC over systems like RSA and DSA known for solving mathematical problem namely ECDLP which takes full exponential time. For this reason, why the ECC offers a high security equal to RSA and DSA while using far smaller key sizes. The attractiveness of ECC will increase relatively to other public-key cryptosystems as the computing power improvements force a general increase within the key size. per-bit include: High speed, Bandwidth savings, Smaller Certificates, Lower power consumption and Storage efficiencies. Trust management between neighbouring vehicles is also challenging due to random distribution and high mobility of vehicles. Trust evaluation is based on numerous factors, such as, recommendations from neighbouring nodes, history of past interactions, reward points etc.

**Related Work:**

In this section, we review various VANET security schemes based on cryptography. Saar et al. [1] have proposed new message authentication ID-MAP (identity-based message authentication using proxy vehicles) scheme based on identity and used proxy vehicles. It ensured message authenticity and provided security against identity attack. Tzeng et al. [2] have presented an efficient IBV (identity-based batch verification) scheme which ensures the privacy preservation and the traceability of vehicles required by the trust authority. Azees et al. [3] proposed an EAAP (efficient anonymous authentication) scheme. It provides a conditional tracking mechanism to trace the vehicles and RSUs that misuse the VANET. Mejri et al. [4] have presented a scheme where group communications make use of a secret key. Only the group members have the knowledge of the secret key; this prevents any intruder to eavesdrop with the group members. Zhu et al. [5] have proposed group signature-based authentication scheme with privacy preservation. This scheme used cooperative authentication, group signature verification, and HMAC. Tangade et al. To encrypt the received message and send a message Pm to B, A chooses a random positive integer k and produces the cipher text Cm as given by equation (1) consisting of the pair of points. Cm= [kG, Pm+kPB] (1) Note that A has used B's public key PB. To decrypt the cipher text, B multiples the first point in the pair by B's private key nB and subtracts the result from the second point as shown by equation (2) PmnB (kG) =Pm+k (nBG)-nB(kG) =Pm (2). The exchange of keys between users A and B can be accomplished as follows: 1. A selects an integer nA less than n. This is A's private key. A then computes and generates a public key PA=nA*G; the public key is a point in Eq(a,b). 2. B similarly selects a private key nB and computes a public key PB, The public keys are exchanged between the nodes A and B. A generates the secret key K=nA*PB. B generates the secret key K=nB*PA. As mentioned in the reference paper [16].

PROPOSED WORK:
SYSTEM ARCHITECTURE:

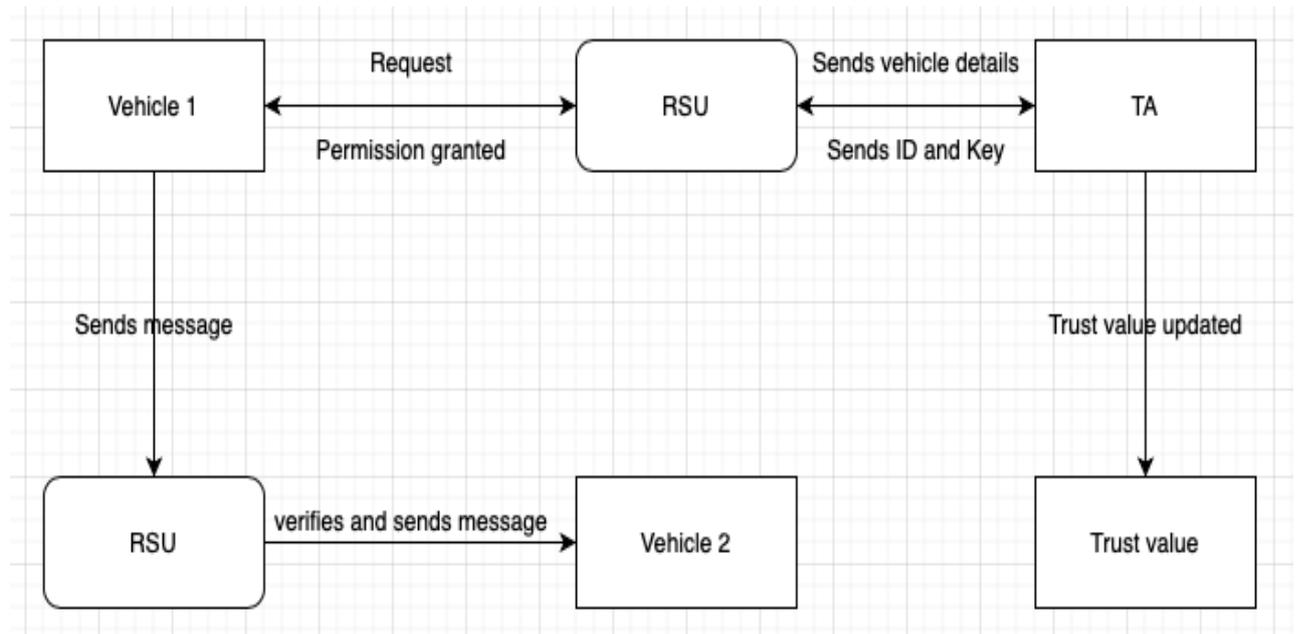

FIG -1

Here, we present a VANET architecture model as shown in Fig. 1. This model consists of four primary components, they are: (i) OBU, (ii) RSU, (iii) Agent of Trusted Authority (ATA) as (Vehicle), and (iv) Regional Transport Office (RTO) as Trusted Authority (TA). The detailed description of these components are as follows. OBUs: It is embedded in each vehicle which enables V2V and V2I communications': These are stationary units deployed on road side for V2I communications. The RSU authenticates and evaluates trust of each vehicle and forwards safety-message to its ATA through wireless communication channel for computation of new trust-value of vehicles. ATA: The Area Post Offices (APOs) are embedded with ATA. It computes new trust-value of vehicles with the help of RSUs and is connected to TA through a secured channel. RTO: It is the supreme trusted entity in VANET called as TA. It manages the off-line registration of ATAs, RSUs, and vehicles (OBU) by generating and issuing private and public keys for these entities. The RSUs are used for authentication purpose with vehicles whereas trust value computation is performed by ATA. While performing trust value computation, RSU and vehicle is not trusted since trust value may be manipulated by them. The RSUs are used for authentication purpose with vehicles whereas trust value computation is performed by ATA. While performing trust value computation, RSU and vehicle is not trusted since trust value may be manipulated by them.

## EXPERIMENTAL SETUP:

Here we used Ns2 simulator and Network animator (NAM) to create the simulation for calculating the trust value.Download and install the ns2 in the PC.Create a TCL(Tool Command Language) file.Set the channel type, radio propagation model, network interface type,MAC type,interface queue type link layer type, antenna model, ax packet in ifq,number of mobile nodes and routing protocol.The number of mobile nodes is set to 31.150 seconds is set as the simulation time.DSDV routing protocol is used for the simulation. Then initialize the ns simulator, trace files and nam files. Configure the nodes and set the nodes position. Trust computation is used to calculate the trust value. Define the DSDV routing protocol. Label and initialize the nodes according to the happenings of the simulation. Run the ns2 simulation in the command terminal using the ns command:

**ns filename.tcl**

The trust levels of each node are shown in the command terminal. The Network animator opens and displays the simulation and the trace files are created. The nam file can be manually opened in command terminal using the ns command

**nam filename.nam**

In the simulation first the cluster node is searched and found.When the malicious node is found it updates the trust level of the node and drops the packets and again the trust level is updated.Again the packets are transmitted between the nodes if any malicious node is found it is eliminated and the trust level is updated.The xgraph can be used to display the bandwidth in transmission of packets between the nodes.It is also used to display the various trust level of nodes at different intervals.

Simulation Parameters

Simulation Area        -  1Km x 1Km

Simulation Time(sec) -  150  secs

Vehicle distribution   -   Random

No of nodes             -  02

Routing protocol       -  DSDV

MAC protocol           -   IEEE 802.11p

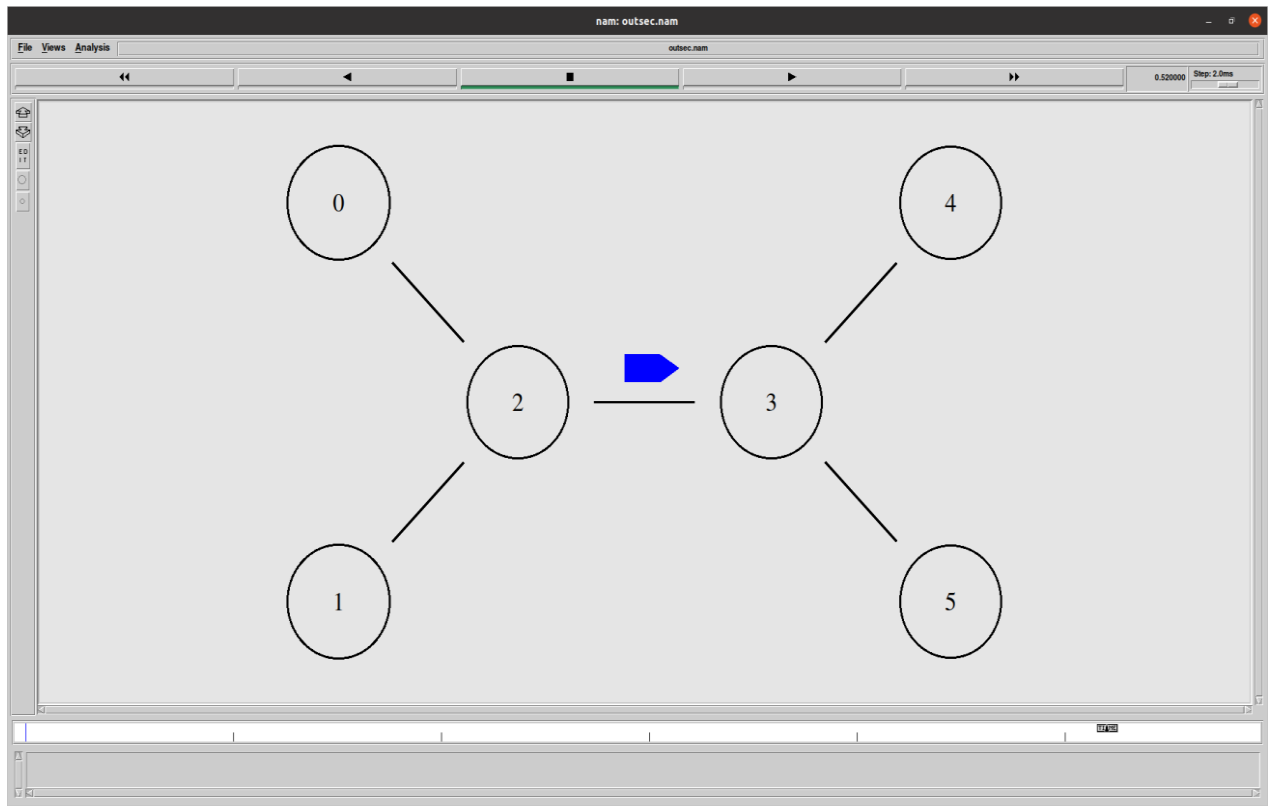

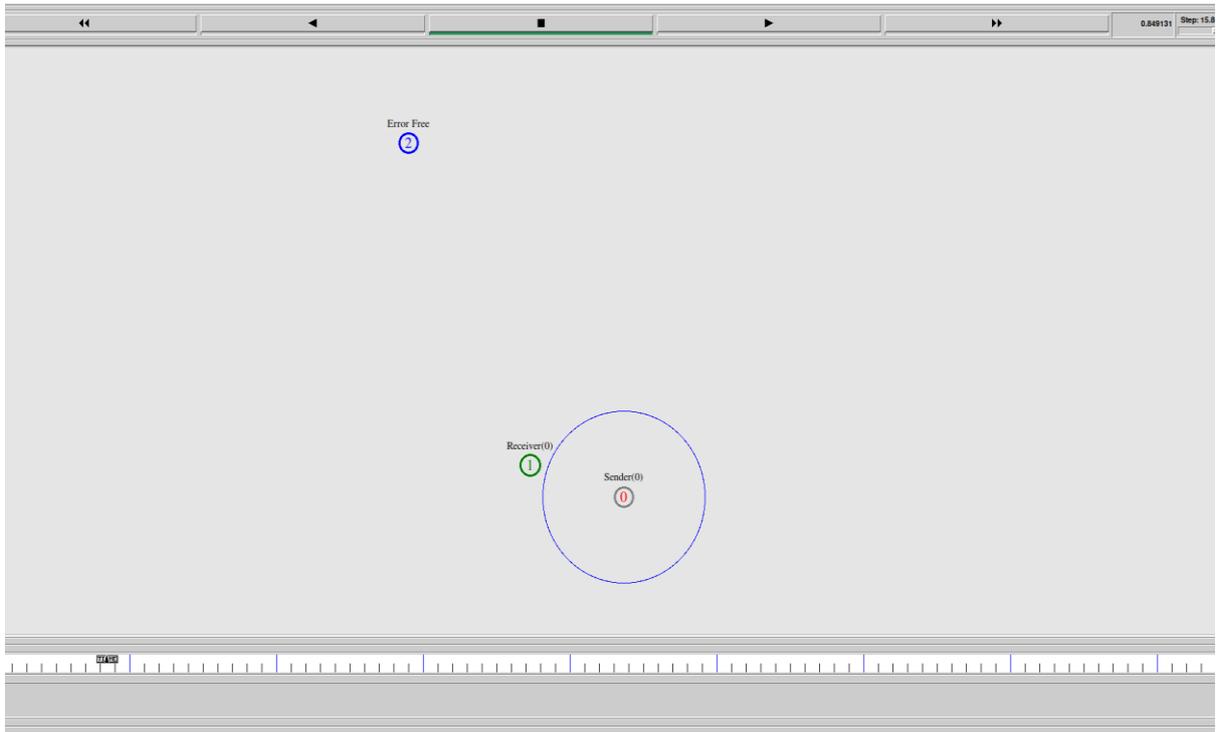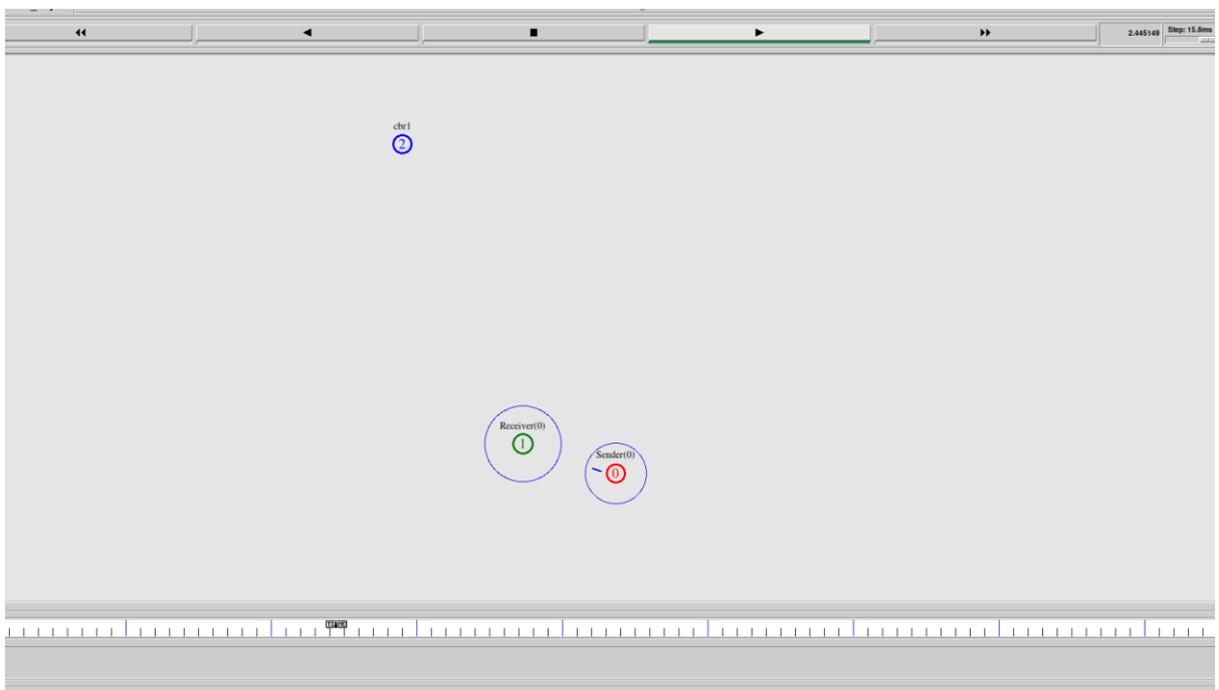

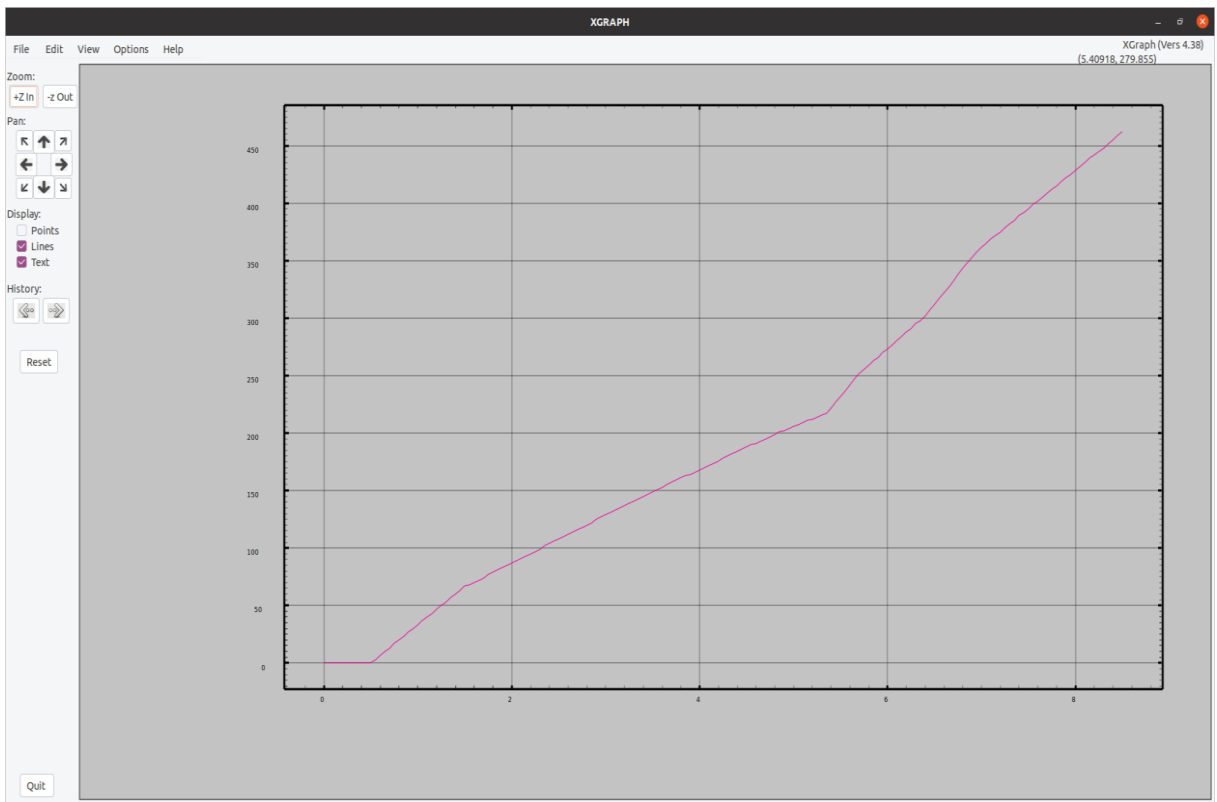

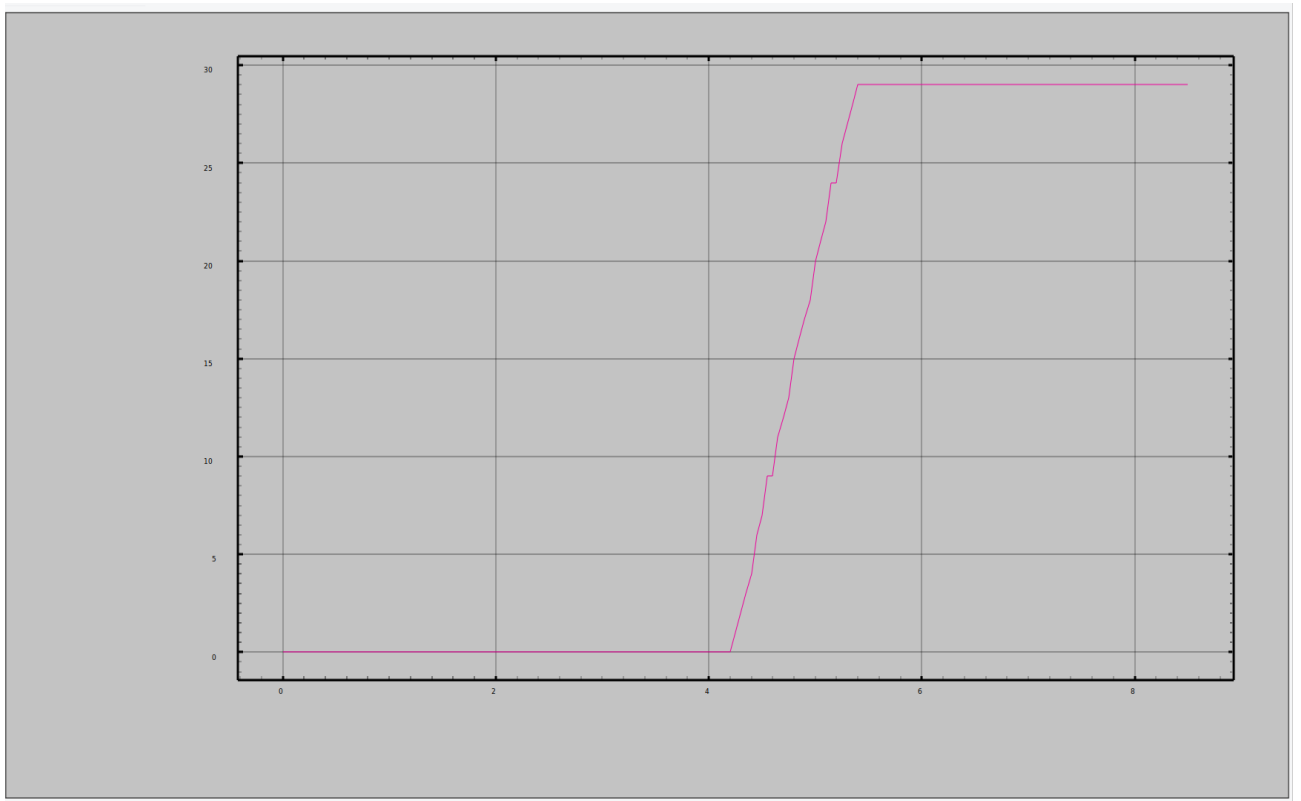

## RESULT AND FUTURE WORK:

In this paper, we have proposed to secure VANET to a larger extent by integrating hybrid cryptography-based authentication and providing safe, secure and efficient delivery of messages. As the results prove a better bandwidth after the delivery of packets to the nodes. As V2I infrastructure would be a challenging task in implementing as for as this model is concerned as collection and transmission of data has a higher complexity level. The proposed scheme is limited to urban areas where RSUs can be deployed. In case of rural areas where RSUs are not deployed we need to consider different model with V2V communications coupled with ATA access through Internet. The performance of proposed scheme can be further improved in terms of speed, computation accuracy by employing batch signature verification or cooperative message verification techniques and adopting machine learning techniques such as reinforcement learning